\documentclass[12pt]{article}
\usepackage{epsf}
\usepackage{amsmath}
\usepackage{amsfonts}
\usepackage{amssymb}
\usepackage{graphicx}
\usepackage{color}
\usepackage{psfrag}
\usepackage{cite}

\usepackage{ifpdf}

\newcommand{\bmat}{\left(\begin{array}}
\newcommand{\emat}{\end{array}\right)}

\def\yzero{\smash{\hbox{$y\kern-4pt\raise1pt\hbox{${}^\circ$}$}}}

\def\beq{\begin{equation}}
\def\eeq{\end{equation}}
\def\beqa{\begin{eqnarray}}
\def\eeqa{\end{eqnarray}}

\def\-{\hphantom{-}}

\def\s2{\frac{1}{\sqrt2}}

\def\beq{\begin{equation}}
\def\eeq{\end{equation}}
\def\beqa{\begin{eqnarray}}
\def\eeqa{\end{eqnarray}}

\def\IF{\relax{\rm I\kern-.18em F}}
\def\II{\relax{\rm I\kern-.18em I}}

\def\Dsl{\,\raise.15ex\hbox{/}\mkern-13.5mu D} 

\def\IS{{\bf S}}
\def\IR{{\bf R}}
\def\IZ{{\bf Z}}
\def\IX{{\bf X}}

\def\IT{{\bf T}}
\def\IP{\bf P}

\def\IM{{\bf M}}

\catcode`\@=11   
\newdimen\@rotdimen
\newbox\@rotbox  

\def\@vspec#1{\special{ps:#1}}
\def\@rotstart#1{\@vspec{gsave currentpoint currentpoint translate
   #1 neg exch neg exch translate}}
\def\@rotfinish{\@vspec{currentpoint grestore moveto}}
\def\@rotr#1{\@rotdimen=\ht#1\advance\@rotdimen by\dp#1%
   \hbox to\@rotdimen{\hskip\ht#1\vbox to\wd#1{\@rotstart{90 rotate}%
   \box#1\vss}\hss}\@rotfinish}
\def\@rotl#1{\@rotdimen=\ht#1\advance\@rotdimen by\dp#1%
   \hbox to\@rotdimen{\vbox to\wd#1{\vskip\wd#1\@rotstart{0 rotate}%
   \box#1\vss}\hss}\@rotfinish}%
\def\@rotu#1{\@rotdimen=\ht#1\advance\@rotdimen by\dp#1%
   \hbox to\wd#1{\hskip\wd#1\vbox to\@rotdimen{\vskip\@rotdimen
   \@rotstart{-1 dup scale}\box#1\vss}\hss}\@rotfinish}%
\def\@rotf#1{\hbox to\wd#1{\hskip\wd#1\@rotstart{-1 1 scale}%
   \box#1\hss}\@rotfinish}%
\def\rotate{\@ifnextchar[{\@rotate}{\@rotate[l]}}
\def\@rotate[#1]#2{\setbox\@rotbox=\hbox{#2}\@nameuse{@rot#1}\@rotbox}

\catcode`\@=12

\topmargin
-1.5cm
\textwidth
15.5cm
\textheight
23.5cm
\oddsidemargin
0.7cm
\evensidemargin
1.2cm

\begin{document}

\makeatletter
\@addtoreset{equation}{section}
\makeatother
\renewcommand{\theequation}{\thesection.\arabic{equation}}
\pagestyle{empty}
\rightline{ IFT-UAM/CSIC-13-105}
\vspace{0.1cm}
\begin{center}
\LARGE{\bf Antisymmetric tensor $\IZ_p$ gauge symmetries \\in field theory and string theory \\[12mm]}
\large{Mikel Berasaluce-Gonz\'alez$^{1,2}$, Guillermo Ram\'{\i}rez$^{1,2}$, Angel M. Uranga$^2$\\[3mm]}
\small{${}^{1}$ Departamento de F\'{\i}sica Te\'orica,\\[-0.3em] 
Universidad Aut\'onoma de Madrid, 28049 Madrid\\
${}^2$ Instituto de F\'{\i}sica Te\'orica IFT-UAM/CSIC,\\[-0.3em] 
C/ Nicol\'as Cabrera 13-15, Universidad Aut\'onoma de Madrid, 28049 Madrid, Spain} \\[2mm] 

\vspace*{2.5cm}

\small{\bf Abstract} \\[5mm]
\end{center}
\begin{center}
\begin{minipage}[h]{16.0cm}
We consider discrete gauge symmetries in $D$ dimensions arising as remnants of broken continuous gauge symmetries carried by general antisymmetric tensor fields, rather than by standard 1-forms. The lagrangian for such a general $\IZ_p$ gauge theory can be described in terms of a $r$-form gauge field made massive by a $(r-1)$-form, or other dual realizations, that we also discuss. The theory contains charged topological defects of different dimensionalities, generalizing the familiar charged particles and strings in $D=4$. We describe realizations in string theory compactifications with torsion cycles, or with background field strength fluxes. We also provide examples of non-abelian discrete groups, for which the group elements are associated with charged objects of different dimensionality.
\end{minipage}
\end{center}
\newpage
\setcounter{page}{1}
\pagestyle{plain}
\renewcommand{\thefootnote}{\arabic{footnote}}
\setcounter{footnote}{0}

\tableofcontents

\vspace*{1cm}

\section{Introduction}

Discrete symmetries are ubiquitous in many models of physics beyond the Standard Model. Moreover, the study of their nature is important also at a more fundamental level, since global symmetries, either continuous or discrete, are believed not to exist in consistent quantum theories including gravity, such as string theory (see the early references \cite{Banks:1988yz,Abbott:1989jw,Coleman:1989zu}, and e.g.\cite{Kallosh:1995hi,Banks:2010zn} for recent discussions). Hence, {\em exact} discrete symmetries should have a gauge nature \cite{Alford:1988sj,Krauss:1988zc,Preskill:1990bm} in these theories.

Discrete gauge symmetries in 4d theories have been subject to  intense study both in field theory (see references above, also \cite{Ibanez:1991hv,Ibanez:1991pr}) and string theory \cite{Camara:2011jg,BerasaluceGonzalez:2011wy, BerasaluceGonzalez:2012vb,BerasaluceGonzalez:2012zn}\footnote{See also \cite{Gukov:1998kn,Burrington:2006uu} and \cite{Ibanez:2012wg,Anastasopoulos:2012zu,Honecker:2013hda,Marchesano:2013ega,Ookouchi:2013gwa} for related applications, and \cite{Kobayashi:2006wq,Lebedev:2007hv,Nilles:2012cy} for discrete symmetries in heterotic orbifolds.}. In these papers,  the discrete symmetries arise as subgroups of continuous gauge symmetries\footnote{Even in cases when there is no obvious underlying continuous symmetry, the latter can be made manifest in suitable supercritical string extensions decaying through closed tachyon condensation \cite{Berasaluce-Gonzalez:2013sna}.}, carried by 1-form fields, broken by their coupling to scalar fields. In a dual formulation
   \cite{Banks:2010zn}, an underlying $\IZ_p$ symmetry is manifest in the existence of a 4d coupling between a 2-form potential $B_2$ and the gauge 1-form (through its field strength)
\beqa
p\int_{4d} B_2\wedge F_2.
\label{pbf}
\eeqa
Since the question of discrete symmetries in theories of quantum gravity is a fundamental one, it is fair to address it in higher-dimensional theories. The latter can host antisymmetric tensor gauge fields of rank higher than those available in 4d. This paper explores a novel realization of discrete gauge symmetries in string theory\footnote{See e.g. \cite{Savit:1977fw,Orland:1981ku} for an early appearance of these gauge symmetries in field theory.}, based on these higher rank gauge fields. In fact, string theory contains a plethora of higher rank antisymmetric tensor fields, which upon compactification pick up topological couplings generalizing (\ref{pbf}). This allows to describe sectors in which two massless antisymmetric tensor gauge fields, of ranks appropriate to the spacetime dimension,  couple and became massive. The gauge symmetry is `broken' by a Higgs-like mechanism, but a discrete $\IZ_p$ subgroup remains.

The main novelty of these $\IZ_p$ gauge symmetries from higher-rank antisymmetric tensors lies in the nature of the charged objects. For discrete symmetries arising from 1-form gauge potentials (coupling to 2-form fields in 4d), the charged objects are $\IZ_p$ particles and $\IZ_p$ strings. The charged objects under these more general $\IZ_p$ symmetries are branes of worldvolume dimension related to the degrees of the form fields involved. Just like particles and strings in 4d, these objects pick up $\IZ_p$ phases when surrounding each other. These objects, and the violation of their number conservation mod $p$, will receive a simple description as suitably wrapped branes in our string theory examples. This description provides an interpretation of the branes characterized by K-theory (or other groups of charges) in terms of topological defects associated to discrete gauge symmetries. 

A complementary interpretation of our results is as a refinement of the discussions in \cite{Quevedo:1996uu} (actually predated by  \cite{Savit:1977fw,Orland:1981ku}), which describe the coupling of two massless antisymmetric tensors of different ranks into one massive antisymmetric tensor. This higher-rank Higgs mechanism motivates a discussion of the phases of the corresponding field theories and their Higgs-confinement dualities. We are thus considering a refinement in which the Higgs-like mechanism leaves an extra $\IZ_p$ discrete gauge symmetry. 

The paper is organized as follows. In Section \ref{sec:field-theory} we provide the field theory description of the phenomenon, starting with the familiar case of 
$\IZ_p$ gauge symmetries from 1-form gauge potentials in section \ref{sec:standard}, and providing the higher-rank generalization in section \ref{sec:higher}. In Section \ref{sec:catalysis} we provide explicit string theory realizations, by exploiting the flux catalysis described in \cite{BerasaluceGonzalez:2012zn} (based on the mechanism in \cite{Maldacena:2001xj}). In Section \ref{sec:torsion}, we comment on a realization in compactifications with torsion homology, generalizing \cite{Camara:2011jg}. In Section \ref{sec:non-abelian}, we discuss the realization of non-abelian discrete symmetries from higher-rank form fields. Finally, Section \ref{sec:conclu} contains some final remarks. Appendix \ref{app:multiple} generalizes the ideas to theories with multiple tensor fields, and clarifies that the discrete symmetries in the original and dual descriptions can be different.

\section{Field theory of higher-rank $\IZ_p$ gauge symmetries}
\label{sec:field-theory}

\subsection{$\IZ_p$ gauge symmetries from 1-form gauge potentials}
\label{sec:standard}

We now quickly review the realization of 4d discrete gauge symmetries as subgroups of `standard' continuous gauge symmetries, i.e. carried by 1-form gauge fields. For simplicity, we stick to the abelian case, which suffices to illustrate the main points. 

The description is phrased in terms of ($a$) a 1-form $A_1$ and a 0-form $\phi$, or ($b$) in a dual version, the magnetic gauge potential $V_1$ and the 2-form $B_2$ \cite{Banks:2010zn} (see also \cite{Hellerman:2010fv} for an alternative viewpoint on discrete gauge symmetries). These are subject to gauge invariances
\beqa
&& (a) \quad A_1\to A_1+d\lambda \quad ,\quad \phi\to \phi + p\,\lambda\nonumber \\
&& (b) \quad B_2\to B_2+d\Lambda_1\quad ,\quad V_1\to V_1+p\,\Lambda_1.
\label{gaugeinv}
\eeqa
This structures lead to a $\IZ_p$ discrete gauge symmetry. The charged objects are $\IZ_p$ particles and strings (electrically charged under $A_1$ and $B_2$, respectively), whose charge can be violated by suitable instantons and junctions, respectively (coupling to $\phi$ and $V_1$, respectively). The processes are associated to the gauge invariant operators \cite{Banks:2010zn}:
\beqa
\exp (-i\phi)\, \exp \Big(\, i\,p\int_LA_1\Big) \quad , \quad  \exp \Big(\, -i\int_C V_1\Big)\, \exp\Big(\, i\,p\int_\Sigma B_2\Big)
\label{standard-ops}
\eeqa
where $L$ is a curve ending at the point $P$ at which $e^{-i\phi}$ is inserted, i.e. $\partial L=P$, and similarly $\Sigma$ is a surface ending on the curve $C$, i.e. $\partial \Sigma=C$.
The first operator in (\ref{standard-ops}) describes $p$ (minimally) charged particles (coupling electrically to $A_1$) along the worldline $L$ emanating from the point $P$; the second describes $p$ (minimally) charged strings (coupling to $B_2$) spanning $\Sigma$ and emanating from a string junction line $C$.

The basic structure of the 4d gauge invariant actions in terms of the above fields is
\beqa
(a)\quad \int_{4d} \, |d\phi -pA_1|^2 \quad \stackrel{\rm dual}{\longleftrightarrow}\quad  (b)\quad \int_{4d} |dV_1-p B_2|^2.
\label{fourd-actions}
\eeqa
In terms of $B_2$ and $A_1$, the $\IZ_p$ discrete symmetry is usually identified from the presence of a 4d topological coupling \cite{Banks:2010zn}
\beqa
p\,\int_{4d} B_2\wedge F_2
\label{pbf2}
\eeqa
between the $U(1)$ field strength $F_2=dA_1$ and the 2-form  $B_2$. We recall that for proper identification of the discrete symmetry, the normalization of $B_2$ is such that its 4d dual scalar has periodicity $1$, and that the minimal $U(1)$ charge is 1.

\subsection{Higher rank $\IZ_p$ discrete gauge symmetries}
\label{sec:higher}

The structure in the previous section is the only one available in four dimensions. However, in higher dimensions there are gauge symmetries carried by higher-rank antisymmetric tensors, and it is reasonable to exploit them to generate discrete $\IZ_p$ gauge symmetries. Conversely, higher dimensions allow the existence of $\IZ_p$ charged objects with higher worldvolume dimensionality. Clearly, a straightforward possibility is to consider a 1-form gauge field and a $(D-2)$-form gauge field in $D$ dimensions, coupling through a $B_{D-2}\wedge F_2$; this is a trivial addition of dimensions, in which the 4d $\IZ_p$ string is extended to a real codimension-2 $(D-3)$-brane, and has appeared implicit or explicitly in earlier discussions of $\IZ_p$ discrete symmetries. In other words, this case can always be dualized into that of a 1-form and a scalar field, i.e. a standard field theory Higgs mechanism.

In this paper we explore $\IZ_p$ symmetries whose underlying continuous symmetry involves genuine higher rank antisymmetric tensors, in any dual picture.
Due to the difficulties with non-abelian tensor field theories, we stick to the abelian case, although Section \ref{sec:non-abelian} contains some discussion on the realization of non-abelian discrete structures.

We consider a theory in $D$ dimensions, with a $r$-form field $A_r$ and a $(r-1)$-form field $\phi_{r-1}$, with the gauge invariance\footnote{These theories have been considered e.g. in \cite{Quevedo:1996uu} (see also e.g. \cite{Troost:1999bj,Diamantini:2001yw}). As in there, we consider the gauge symmetries to be {\em compact}, namely there is charge quantization for the extended objects to which they couple. As in the 4d case, normalization is such that the minimal charge is unity.}
\beqa
A_r\to A_r+d\lambda_{r-1} \quad ,\quad \phi_{r-1}\to \phi _{r-1}+ p\,\lambda_{r-1}.
\label{gauge-inv-gen1}
\eeqa 
The notation is obviously chosen to recover the familiar one for $r=1$, c.f. (\ref{gaugeinv}a). A gauge invariant action, generalizing (\ref{fourd-actions}a), is
\beqa
\int_{\IM_D} |\, d\phi_{r-1} - p A_r \,|^2.
\label{action-gen1}
\eeqa
Notice that in this theory both fields are gauge fields since, on top of (\ref{gauge-inv-gen1}), the lagrangian is invariant under $\phi_{r-1}\to \phi_{r-1}+d\sigma_{r-2}$.

In the above lagrangian, the field $A_r$ eats up the field $\phi_{r-1}$ and gets massive. Note that this is consistent with the counting of degrees of freedom of antisymmetric tensor gauge fields under the $SO(D-2)$ and $SO(D-1)$ little groups for massless and massive particles:
\beqa
 \begin{pmatrix} D-2 \cr r \end{pmatrix} + \begin{pmatrix} D-2 \cr r-1 \end{pmatrix} =  \begin{pmatrix} D-1 \cr r \end{pmatrix} .
\eeqa
The gauge symmetry of $A_r$ is broken spontaneously, but a discrete $\IZ_p$ symmetry remains. This is a higher-rank analogue of the Higgsing of a $U(1)$ gauge group by eating up the phase of a charge-$p$ scalar.
However, the naturally charged objects are not in general codimension-2 $(D-3)$-branes, and point particles, as in the rank-1 case; rather we have $(r-1)$-branes and $(D-r-2)$-branes (electric charges under $A_r$ and magnetic charges under $\phi_{r-1}$). 

It is straightforward to dualize $A_r$ into its magnetic $(D-r-2)$-form gauge potential $V_{D-r-2}$, and $\phi_{r-1}$ into its dual $(D-r-1)$-form gauge potential $B_{D-r-1}$. They are subject to the gauge invariance c.f. (\ref{gaugeinv}b)
\beqa
B_{D-r-1}\, \to \, B_{D-r-1}\, +\, d\Lambda_{D-r-2} \quad , \quad V_{D-r-2}\, \to\,  V_{D-r-2}\, +\, p\, \Lambda_{D-r-2}.
\label{gauge-inv-gen2}
\eeqa
The dual gauge-invariant action has the structure
\beqa
\int_{\IM_D} \, |dV_{D-r-2}\, -\, p\, B_{D-r-1}|^2.
\label{action-gen2}
\eeqa
This dual description makes manifest an {\em emergent} $\IZ_p$ gauge symmetry\footnote{An important point, not manifest in the examples in \cite{Banks:2010zn} is that the emergent discrete symmetry may differ from the original one. This is illustrated explicitly in Appendix \ref{app:multiple}.}. The objects charged under $\phi_{r-1}$ and $V_{D-r-2}$ are $(r-2)$- and $(D-r-3)$-branes, and play the role of generalized junctions violating the number of $(r-1)$- and $(D-r-2)$-branes in $p$ units, making them $\IZ_p$-valued. This follows form the gauge-invariant operators
\beqa
\exp\Big(\!- i \!\!\int_{P_{r-1}}\!\!\! \phi_{r-1} \Big) \, \exp \Big(\, i\, p\, \int_{L_{r}}\!\! A_r \Big)\quad , \quad
\exp\Big(\! -i \!\!\int_{C_{D-r-2}} \!\!\!\!\!\!V_{D-r-2} \Big) \, \exp \Big(\, i\, p\, \int_{\Sigma_{D-r-1}}\!\!\!\!\!\! B_{D-r-1} \Big)\nonumber
\eeqa 
where $L_r$ has $P_{r-1}$ as its boundary, $\partial L_r=P_{r-1}$, and similarly $\partial\Sigma_{D-r-1}=C_{D-r-2}$.

By standard arguments, the quantum amplitude of a process involving a (minimally charged) $(r-1)$-branes with worldvolume $\Sigma_r$, and a (minimally charged) $(D-r-2)$-brane with worldvolume $\Delta_{D-r-1}$ receives a phase 
\beqa
\exp \bigg[\,\frac{2\pi i}{p} \, L(\Sigma_r,\Delta_{D-r-1})\, \bigg]
\eeqa
where $L(\Sigma_r,\Delta_{D-r-1})$ is the linking number in $D$-dimensions (the number of times $\Sigma_r$ surrounds $\Delta_{D-r-1}$, or vice-versa). 

For future convenience, it is useful to identify the analogue of the $BF$ coupling (\ref{pbf2}) in 4d. This is the topological coupling 
\beqa
p\,\int_{\IM_D} B_{D-r-1}\wedge F_{r+1}\, ,
\label{higher-BF}
\eeqa
where we have introduced the field strength $F_{r+1}=dA_r$.

The construction in this section is  basically a refinement of that in \cite{Quevedo:1996uu}. The main novelty is the identification of the unbroken $\IZ_p$ symmetry, which reflects in a $\IZ_p$-grading of the topological defects in the theories under consideration. In these $\IZ_p$ theories, the duality between the Higgs and confinement phases holds as in  \cite{Quevedo:1996uu}. Indeed, the Higgs phase of $A_r$ translates into the fact that the dual magnetic $(D-r-3)$-branes cannot exist in isolation but are confined by the $p$ $(D-r-2)$-branes stuck to them.

\section{Higher-rank $\IZ_p$ symmetries in string theory flux compactifications}
\label{sec:catalysis}

A simple way to realize rich sets of $\IZ_p$-charged objects, associated to discrete gauge symmetries, is the `flux catalysis' systematically studied in \cite{BerasaluceGonzalez:2012zn} (based on  \cite{Maldacena:2001xj}) for 4d discrete symmetries carried by 1-form gauge fields; see \cite{Evslin:2001cj,Evslin:2002sa,Evslin:2003hd,Evslin:2004vs,Collinucci:2006ug,Evslin:2007ti,Evslin:2007au} for related phenomena. The key idea is that in 4d string compactifications with field-strength flux backgrounds (flux compactifications), the 10d Chern-Simons couplings can produce 4d $BF$ couplings associated to $U(1)$ gauge symmetries broken to $\IZ_p$ subgroups.

Clearly, the idea easily generalizes to produce $\IZ_p$ gauge symmetries from higher rank antisymmetric tensor fields. In this Section we pursue this suggestion to recover the structures introduces in section \ref{sec:higher}, for compactifications to higher $D>4$.

\medskip

For concreteness, we focus on a particular example in $D=6$. Consider a compactification of type IIA on a real dimension 4 space $\IX_4$ (not necessarily $\IT^4$ or K3, since we are not particularly interested in supersymmetry). We introduce $p$ units of flux for the RR field strength 4-form $F_4=dC_3$
\beqa
\int_{\IX_4} F_4\, =\, p.
\eeqa
The 10d Chern-Simons couplings produce the following 6d coupling
\beqa
\int_{10d} \, C_3\wedge H_3\wedge F_4 \; \rightarrow \; p \int_{6d} C_3\wedge H_3.
\eeqa
This has the structure (\ref{higher-BF}) for $r=2$ (with $B_3\to C_3$ and $F_3\to H_3$). There are $\IZ_p$-charged 1-branes (arising from fundamental F1-strings) and 2-branes (from D2-branes). Their decay occurs through junctions of worldvolume dimensions 1 and 2, respectively, from the corresponding dual sources; namely, a D4-brane wrapped on $\IX_4$ (on which $p$ F1-strings must end \cite{Witten:1998xy}), and a NS5-brane wrapped on $\IX_4$ (on which $p$ D2-branes must end, by a dual of the Freed-Witten anomaly\footnote{Actually \cite{Freed:1999vc} considered the case of torsion $H_3$ flux, and the physical picture for general $H_3$ appeared in \cite{Maldacena:2001xj}. Still, we stick to the widely used term Freed-Witten anomaly, even for non-torsion fluxes.} \cite{Maldacena:2001xj}). See Appendix B of \cite{BerasaluceGonzalez:2012zn} for an overview of these processes.

The M-theory version of the above system is interesting, and arises naturally in the context of the AdS$_7$/CFT$_4$ correspondence. Compactification of M-theory on a 4-manifold down to $D=7$, with $p$ units of $G_4$ 4-form flux produces a 7d coupling $p\int_{7d} G_4\wedge C_3$. The corresponding $\IZ_p$ discrete symmetry has appeared in \cite{Aharony:1998qu}. It is amusing to notice that M2-branes correspond to the two kinds of $\IZ_p$ topological defects, hence M2-branes pick up $\IZ_p$ phases when surrounding each other, in a higher dimensional analogy of anyons in $D=3$. This interesting behaviour is presumably linked to the elusive system of coincident M5-branes underlying this gauge/gravity duality.

\section{Higher-rank $\IZ_p$ symmetries in string compactifications with torsion}
\label{sec:torsion}

Discrete gauge symmetries associated to higher-rank forms are briefly mentioned in \cite{Moore:1999gb}, although related to torsion in homology or K-theory. This very formal discussion can be made very explicit following \cite{Camara:2011jg}, at least for torsion homology. To show that compactifications with torsion homology can produce higher-rank discrete symmetries, we consider a simple illustrative example. Consider M-theory on  a 4-manifold with torsion 1-cycles (and their dual 2-cycles), $H_1(\IX_4,\IZ)=H_2(\IX_4,\IZ)=\IZ_p$. We focus on the sector of M2-branes on 1-cycles -- 7d strings -- and M5-branes on 2-cycles -- 7d 3-branes -- (there is another sector of M2-branes on 2-cycles and M5-branes on 1-cycles, which can be discussed similarly). Following \cite{Camara:2011jg}, we introduce the Poincar\'e dual torsion 2- and 3-forms $\alpha_2^{\rm tor}$, ${\tilde \omega}_3^{\rm tor}$, satisfying the relations
\beqa
d\omega_1^{\rm tor}=p\, \alpha_2^{\rm tor} \quad , \quad d\beta_2^{\rm tor}=p\, {\tilde \omega}_3^{\rm tor}
\eeqa
where $\omega_1^{\rm tor}$ and $\beta_2^{\rm tor}$ are globally well-defined 1- and 2-forms. The torsion 2- and 3-forms $\alpha_2^{\rm tor} $ and ${\tilde \omega}_3^{\rm tor}$ are thus trivial in de Rham cohomology, but not in the $\IZ$-valued cohomology, i.e. $H_2(\IX_4\IR)=H_3(\IX_4,\IR)=\emptyset$, $H_2(\IX_4\IZ)=H_3(\IX_4,\IZ)=\IZ_p$. The torsion linking number is encoded in the intersection pairing
\beqa
\int_{\IX_4} \alpha_2^{\rm tor}\wedge \beta_2^{\rm tor}=\int_{\IX_4} \omega_1^{\rm tor}\wedge {\tilde \omega}_2^{\rm tor}=1.
\eeqa
These forms are assumed to be eigenstates of the Laplacian \cite{Camara:2011jg}, corresponding to massive modes; they can be usefully exploited to describe dimensional reduction of the antisymmetric tensor fields, in particular, the M-theory 3- and 6-forms
\beqa
C_3= \phi_1 \wedge \alpha_2^{\rm tor}\, +\, A_2\wedge \omega_1^{\rm tor}
\quad ,\quad 
C_6= B_4\wedge \beta_2^{\rm tor} \, +\, V_3\wedge {\tilde \omega}_3^{\rm tor}.
\eeqa
The corresponding field strengths contain the structures
\beqa
dC_3 =(d\phi_1+pA_2)\wedge \alpha_2^{\rm tor}+ \ldots \quad , \quad dC_6 = (dV3+pB_4)\wedge {\tilde \omega}_3^{\rm tor} +\ldots
\eeqa
which (modulo a trivial sign redefinition) imply the gauge invariances (\ref{gauge-inv-gen1}), (\ref{gauge-inv-gen2}). Accordingly, the 11d kinetic term for $G_4=dC_3$ (and its dual) lead to 7d actions with the structure (\ref{action-gen1}), (\ref{action-gen2}). The dimensional reduction we have just sketched thus relates the underlying torsion homology with the $\IZ_p$ gauge theory lagrangians of section \ref{sec:higher}.

\section{The non-abelian case}
\label{sec:non-abelian}

Non-abelian discrete gauge symmetries are interesting\footnote{See \cite{Alford:1989ch,Alford:1990mk,Alford:1990pt,Alford:1991vr,Alford:1992yx,Lee:1994qg} for early field theory literature, and \cite{Gukov:1998kn,BerasaluceGonzalez:2012vb,BerasaluceGonzalez:2012zn} for string realizations in type II and \cite{Kobayashi:2006wq,Nilles:2012cy} in heterotic orbifolds.}. In 4d, the non-abelian character can be detected by letting two strings (with charges given by non-commuting group elements $a$, $b$) cross, and watching the appearance of an stretched string (with charge given by the commutator $c=aba^{-1}b^{-1}$). In string theory realizations, this follows from brane creation processes when the underlying branes are crossed \cite{Hanany:1996ie}.

In general dimension $D$, we can look for similar effects, the only difference being that the objects have richer dimensionality. Consider the following table, which describes the geometry of two branes (denoted 1 and 2) which cross and lead to the creation of brane 3
\begin{center}
\begin{tabular}{ccccc}
Brane 1 & $\overbrace{- \cdots -}^{d_1}$ &  $\overbrace{- \cdots -}^{d_2}$ &  $\overbrace{\times\cdots \times}^{d_3}$ & $\times$\\
Brane 2 & $-\cdots -$ & $\times \cdots\times$ & $-\cdots -$ & $\times$ \\
Brane 3 & $-\cdots - $ & $\times \cdots\times$ & $\times \cdots\times$ & $-$
\end{tabular}
\end{center}
The symbols $-$ and $\times$ denote that the brane spans or does not span the corresponding dimension, and obviously $d_1+d_2+d_3+1=D$. The last entry corresponds to the single overall transverse dimensions to branes 1 and 2, on which the crossing proceeds, and along which the created brane 3 stretches.

As a concrete example, involving discrete gauge symmetries arising from torsion homology c.f. section \ref{sec:torsion}, consider type IIB compactified on a 5-manifold with a $\IZ_p$ torsion 3-cycle, self-intersecting over the dual $\IZ_p$ torsion 1-cycle (the AdS$_5\times \IS^5/\IZ_3$ geometry in \cite{Gukov:1998kn} is a realization for $p=3$). The theory contains 5d 2-branes arising from NS5-branes on the torsion 3-cycle, a further set of 5d 2-branes from D5-branes on the torsion 3-cycle, and a set of 5d 2-branes from D3-branes on the torsion 1-cycle. The crossing of NS5- and D5-branes produces D3-branes \cite{Hanany:1996ie}, leading to the above 2-brane crossing effect (with $d_1=2$, $d_2=d_3=1$ in the above table); the resulting discrete group is non-abelian, and is given by a $\Delta_{27}$ (for general $\IZ_p$ torsion, a discrete Heisenberg group \cite{BerasaluceGonzalez:2012vb}, see also \cite{Burrington:2006uu}).

One can similarly construct more exotic examples, in which the non-abelian symmetry group elements are associated to objects of {\em different} dimensionality. For instance, consider type IIA compactified on  the same geometry as above, i.e. a 5-manifold with torsion 3- and 1-cycles.  The theory contains 5d 2-branes from NS5-branes on the torsion 3-cycle, a set of 5d 1-branes from D4-branes on the torsion 3-cycle, and a further set of 5d 1-branes from D2-branes on the torsion 1-cycle. The crossing of NS5- and D4-branes produces D2-branes; in 5d the process corresponds to crossing a 2-brane with a 1-brane, with the creation of another kind of 1-brane (hence we have $d_1=1$, $d_2=2$, $d_3=1$). The resulting discrete Heisenberg symmetry group is exotic, since its elements are associated to objects of different dimensionality. A similar phenomenon already occurs in the (abelian) context of D-brane charge classification by K-theory, where in certain examples the charges in a K-theory group  correspond to  branes in cohomology classes of different degree (e.g. \cite{Moore:1999gb} quotes the example of $\IR\IP_7$, where the torsion cohomology is $\IZ_2\oplus\IZ_2\oplus\IZ_2$, with the torsion K-theory is $\IZ_8$).

Similar examples could be worked out involving branes  whose charges are $\IZ$-valued in (co)homology, but which are actually torsion due to the presence of background fluxes.
We refrain from a systematic discussion, hoping that the above examples suffice to illustrate the main idea.

\section{Final remarks}
\label{sec:conclu}

In this paper we have considered discrete gauge symmetries remaining from broken continuous gauge symmetries carried by general antisymmetric tensor fields. We have described the field theory for these general $\IZ_p$ gauge theories, in several dual realizations\footnote{In fact, we have described in appendix \ref{app:multiple} that the discrete gauge symmetry can in general change in the dual description, in the presence of additional continuous gauge symmetries.}. We have described abelian and non-abelian realizations in string theory, in compactifications with torsion cycles, or generating torsion by flux catalysis. We would like to conclude with a few remarks:

$\bullet$ The case of 1-form gauge symmetries broken by scalars can be elegantly described in the language of gaugings in supergravity. It would be interesting to develop such a description for the higher-rank case. 

$\bullet$ The non-abelian structure of Section \ref{sec:non-abelian} is intriguing, as it points to some underlying non-abelian (broken) symmetry involving higher-rank antisymmetric tensors (possibly of different degree). It would be interesting to explore the existence of this underlying structure more directly, possibly in terms of non-abelian gaugings.

$\bullet$ Recent holographic discussions of the gravitational dual to certain superconductors (e.g. helical or striped phase $p$-wave superconductors \cite{Donos:2012gg}) involve 
2-form fields with topological couplings to 1-form gauge fields in $D=5$. It would be interesting to explore possible holographic applications of our higher-rank antisymmetric tensor field theories, and their discrete symmetries.  

We hope this paper triggers further progress into understanding discrete gauge symmetries, and the role of higher-rank gauge potentials, in field theory and string theory.

\bigskip

\centerline{\bf \large Acknowledgments}

\bigskip

We thank P. Orland for pointing out early references on $\IZ_p$ gauge field theories carried by higher rank antisymmetric tensors.
This work has been supported by the Spanish Ministry of Economy and Competitiveness under grants FPA2010-20807, FPA2012-32828, Consolider-CPAN (CSD2007-00042),  and SEV-2012-0249 of the Centro de Excelencia Severo Ochoa Programme,  by the  Comunidad de Madrid under grant HEPHACOS-S2009/ESP1473, and by the  European Commission under contract ERC-2012-ADG$\_$20120216-320421 (SPLE Advanced Grant)  and PITN-GA-2009-237920 (UNILHC network). M.B-G. acknowledges the finantial support of the FPU grant AP2009-0327. G.R. acknowledges the support of Campus Excelencia Internacional UAM+CSIC.

\newpage

\appendix

\section{Multiple antisymmetric tensors}
\label{app:multiple}

In this section we comment on subtle points arising when the topological couplings between antisymmetric tensor gauge fields involve several fields of each kind. The analysis is similar to section \ref{sec:higher}, with additional subtleties in identifying the emergent discrete gauge symmetry in the dual description (using ingredients partially noticed in \cite{BerasaluceGonzalez:2012zn}).

\subsection{Field theory description}

For instance, consider a single $r$-form gauge field $A_r$ made massive by coupling to several $(r-1)$-form fields $\phi_{r-1}^k$ in $D$ dimensions, with the lagrangian
\beqa
\int_{D} \sum_k |d\phi_{r-1}^k \,-\, p_k\, A_r|^2.
\label{multi-one}
\eeqa
This is  gauge invariant under
\beqa
A_r & \to & A_r+d\lambda_{r-1} \nonumber\\
\phi_{r-1}^k & \to & \phi_{r-1}^k \, +\, p_k\, \lambda_{r-1}.
\eeqa
The potential $A_r$ actually eats up only one linear combination of the fields $\phi_{r-1}^k$, while the orthogonal linear combinations remain as massless $(r-1)$-form fields. Denoting $p={\rm g.c.d}(p_k)$, the massive gauge symmetry leaves a remnant $\IZ_p$ gauge symmetry. This follows from the structure of $\IZ_p$ charged $(r-1)$-brane states, whose number can be violated by operators
\beqa
\exp \bigg(-i\int_{P_{r-1}}\phi_{r-1}^k\bigg) \,\exp \bigg(i\int_{L_r} p_kA_r\bigg).
\eeqa
Each such vertex creates $p_k$ $(r-1)$-branes, so by Bezout's lemma, there exists a set of vertices which  (minimally) violates their number in  $p$ units, making the $(r-1)$-branes $\IZ_p$-valued. In addition, the theory enjoys the continuous gauge invariance associated to the orthogonal combinations of the $\phi_{r-1}^k$'s.

In the dual realization, we have a single potential $V_{D-r-2}$ and several potentials $B_{D-r-1}^k$, with lagrangian
\beqa
\int_{D} |dV_{D-r-2}\, -\, {\textstyle\sum_k }\,p_k \, B_{D-r-1}^k|^2.
\eeqa
There are gauge invariances under
\beqa
B_{D-r-1}^k &\to & B_{D-r-1}^k\, +\, d\Lambda_{D-r-2}^k\nonumber \\
V_{D-r-2} & \to & V_{D-r-2}\, +\, {\textstyle \sum_k} p_k\Lambda_{D-r-2}^k
\eeqa
(on top of the dual gauge transformation $V_{D-r-2}\to V_{D-r-2}+d\sigma_{D-r-3}$). One combination of the continuous gauge symmetries, given by $\sum_k (p_k/p) T_k$ (where $T_k$ is the generator of the $k^{th}$ gauge transformation), is actually broken to a discrete subgroup $\IZ_q$, with $q=\sum_k (p_k)^2/p$ \cite{BerasaluceGonzalez:2012zn}. Hence, the discrete part of the emergent gauge group in the dual description is different from the original one; this is a novel feature as compared with the system in \cite{Banks:2010zn} and in Section \ref{sec:field-theory}. The $\IZ_q$ structure follows from the structure of charged $(D-r-2)$-brane states, which are created by operators
\beqa
\exp\bigg( -i\int_{C_{D-r-2}} V_{D-r-2}\bigg)\, \exp \bigg( i\int_{\Sigma_{D-r-1}} \sum_k \, p_k\, B_{D-r-1}^k \bigg).
\eeqa
This violates $T_k$ charge conservation in $p_k$ units, and hence $\sum_k (p_k/p) T_k$ in $q=\sum_k (p_k)^2/p$ units. 

The fact that the original $\IZ_p$ and the emergent $\IZ_q$ gauge symmetries are different is not in contradiction with charge quantization of the dual charged objects, i.e. the $\IZ_p$ $(r-2)$-branes and the $\IZ_q$ $(D-r-2)$-branes, because of the presence of additional charges under the additional continuous gauge symmetries in the system. 

Clearly, a similar (but more involved) analysis can be carried out when there are several fields of each kind. We leave this for the interested reader.

\subsection{A string theory example}

It is easy to use e.g. the flux catalysis of Section \ref{sec:catalysis} to obtain concrete examples of the above structure. For instance, we consider the example of type IIA compactified on K3 to $D=6$, with background $F_2$ flux. Specifically, we introduce two basis of 2-cycles  $\{\alpha_k\}$, $\{\beta_k\}$, with $\alpha_k\cdot\beta_l=\delta_{kl}$, and define
\begin{equation}
\int_{\alpha_k}F_2=p_k.
\end{equation}
There a 6d topological coupling arising as follows
\begin{eqnarray}
\int_{10d} B_2\wedge F_2\wedge F_6 \quad\longrightarrow\quad\sum_k\int_{6d}p_k\, B_2\wedge\hat{F}_4^k
\label{K3IIAF2b}
\end{eqnarray}
where
\begin{equation}
\hat{F}_4^k=\int_{\beta_k}F_6.
\end{equation}
This mixed term has the structure to complete into the square
\beqa
\int_{6d} |  d\phi_1^k-p_k B_2|^2
\eeqa
where we have introduced the 6d duals of of $\hat{C}_3^k$, given by $\phi_1^k=\int_{\alpha_k} C_3$. This has the structure (\ref{multi-one}) with a trivial notation change.

\end{document}